\documentstyle[twocolumn,prl,aps,epsfig]{revtex}

\begin{document}
\title{Decoherence in circuits of small Josephson junctions}
\author{J.P. Pekola and J.J. Toppari}
\address{Department of Physics, University of Jyv\"askyl\"a, P.O. Box 
35 (Y5), FIN-40351, Jyv\"askyl\"a, Finland \vspace{16mm}}
\maketitle

\begin{minipage}{170mm}
\vspace{-31mm}
\begin{quote}
\begin{abstract}
We discuss dephasing by the dissipative electromagnetic environment 
and by measurement in circuits consisting of small Josephson 
junctions. We present quantitative estimates and determine in which 
case the circuit might qualify as a quantum bit. Specifically, we 
analyse a three junction Cooper pair pump and propose a measurement to 
determine the decoherence time $\tau_\varphi$.
\end{abstract}
\end{quote}
\end{minipage}
\vskip -4mm

The rate at which phase coherence is lost is an important factor in 
deciding whether a system qualifies as a quantum bit. Until now there 
exist very few experimental tests \cite{nakamura} or theoretical 
arguments \cite{averin,schon} to determine this time in proposed 
quantum bits based on small Josephson junctions (squbits), and 
moreover only a lower bound of $\sim 5$ ns has been determined so far. 
In this letter we estimate quantitatively the dephasing time 
$\tau_\varphi$ caused by the electromagnetic environment in 
single- and multijunction Josephson circuits both at zero and at 
non-zero temperatures. This dephasing is induced by the same 
fundamental processes which cause the decoherence in a squbit, thus 
yielding a direct measure of the decoherence time of a squbit 
\cite{cottet}. Based on our analysis we determine certain 
limiting factors for the realisation of a successful squbit 
experiment, and as a concrete example we investigate the three 
junction Cooper pair pump in which the coherent nature of the charge 
transport induces deviations from the accurate quantized transfer 
\cite{omapump}. We demonstrate that the crossover at $f=f_C$ as a 
function of the  pumping frequency $f$ from incoherent to coherent 
charge transport yields a direct measure of $\tau_\varphi \simeq 
1/f_C$. We also discuss the limitations in implementing the so-called 
quasiparticle traps often used to suppress single electron effects in 
Cooper pair transistors and Cooper pair boxes \cite{joyez}.  

We start by relating the dephasing rate of the Josephson junction 
circuit under consideration to the impedance of the electromagnetic 
environment that it is imbedded in. By dephasing we mean the deviation 
of the Josephson phase $\varphi(t)$ across the junction circuit from 
its initial value $\varphi(0)$. Specifically we are interested in the 
rms value of the phase deviation 
$\sqrt{\langle(\Delta\varphi)^2\rangle} \equiv \sqrt{ 
\langle[\varphi(t)-\varphi(0)]^2\rangle}$. Voltage fluctuations 
induced by dissipative circuit elements result in the phase-phase 
correlation function $J(t) = 
\langle\left[\varphi(t)-\varphi(0)\right]\varphi(0)\rangle$ 
\cite{grabert,nazarov}, which, based on the fluctuation-dissipation 
theorem, can be expressed in the form
\begin{eqnarray}
\label{jt}
J(t)&=&2\int\limits_0^\infty\frac{d\omega}{\omega}\frac{{\rm 
Re}Z_t(\omega)}{R_K}\times\\
&&\left\{ 
\coth\!\!\left(\frac{\hbar\omega}{2k_BT}\right)\!\left[\cos(\omega 
t)\!-\!1\right] -i\sin(\omega t) \right\}\nonumber,
\end{eqnarray}  
where $Z_t(\omega)$ is the impedance seen by the circuit/junction 
whose phase fluctuations we want to determine, and $R_K = h/e^2 \simeq 
25.8$ k$\Omega$ is the resistance quantum. It is straightforward to 
see that $\langle(\Delta\varphi)^2\rangle$ and $J(t)$ are related by 
\begin{equation}
\left\langle\left(\Delta\varphi\right)^2\right\rangle = -2{\rm 
Re}J(t).
\end{equation}
Other possible sources of decoherence \cite{averin} besides the 
dissipative environment will not be discussed here. 


\begin{figure}[ht]
\center
\epsfig{file=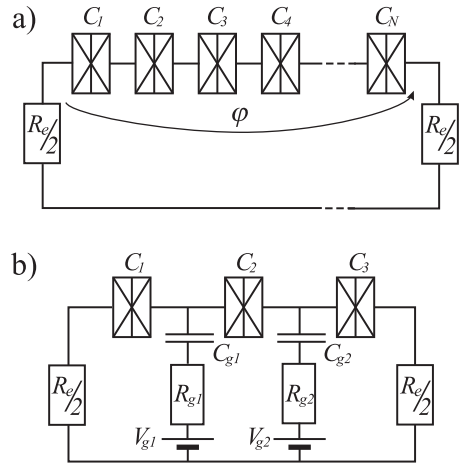,width=60truemm}
\vskip 2mm
\caption{Schematic views of two circuits studied quantitatively. {\bf 
a)} An $N$-junction array with purely resistive electromagnetic 
environment. {\bf b)} A three junction Cooper pair pump with a purely 
resistive environment in biasing and in gate lines.}
\label{schem}
\end{figure}
First we consider the case of an array of $N$ superconducting tunnel 
junctions. In the following analysis the array is assumed to be 
homogeneous, $C_1 = C_2 = \cdots = C_N \equiv C$, and the 
electromagnetic environment to be purely resistive $Z(\omega) = R_e$ 
as shown in Fig.~\ref{schem}a. (Figure \ref{schem}b refers to the 
three junction Cooper pair pump to be discussed later.) These 
simplifications make results more transparent, but it is 
straightforward to generalise the present method also to other 
circuits with an arbitrary environment, $Z(\omega)$. With the 
assumptions mentioned we obtain for the total resistance seen by the 
array \cite{grabert,nazarov} $Z_t(\omega) = R_e/(1+i\omega\tau)$, 
where $\tau=R_eC/N$. The real part of this can be written in form
\begin{equation}
\label{ztr}
{\rm Re}Z_t(\omega)=\frac{R_e}{1+\omega^2\tau^2}\,.
\end{equation}
Inserting this into Eq.~(\ref{jt}) and using the result for $J(t)$ 
when Re$Z_t(\omega)$ assumes the Lorentzian form of Eq.~(\ref{ztr}) 
\cite{jt,jt2} we obtain an expression for 
$\langle(\Delta\varphi)^2\rangle$. In the limit of zero temperature 
($T\rightarrow 0$) we immediately get for $t\gg\tau$
\begin{equation}
\left\langle\left(\Delta\varphi\right)^2\right\rangle = 
4\frac{R_e}{R_K}\left[\,\ln\!\left(t/\tau\right)+\gamma \right],
\label{T0}
\end{equation}
where $\gamma \simeq 0.57721$ is Euler's constant. In the case of 
non-zero temperature we consider only the long time ($\pi 
k_BTt/\hbar\gg 1$) limit which is relevant in most cases, except in 
the limit of large $R_e$. At a realistic measurement temperature, e.g. 
$T= 50$ mK, the result is valid in the range $t\gg 50$ ps, which is 
the region we are interested in. Long time expansion yields
\begin{equation}
\left\langle\left(\Delta\varphi\right)^2\right\rangle \simeq 
4\frac{R_e}{R_K} \left[\frac{\pi k_BT}{\hbar}t - 
\ln\!\!\left(\frac{2\pi k_BT\tau}{\hbar}\right) + \gamma\right].
\label{lt}
\end{equation}
The long time expansion is valid only at non-zero temperatures and 
therefore Eq.~(\ref{T0}) cannot be recovered from Eq.~(\ref{lt}) in 
the limit of $T\rightarrow 0$.   

We can also apply the same method to an individual (the $i$th) 
junction to find the phase fluctuations 
$\langle(\Delta\varphi_i)^2\rangle$ across it. With the same 
assumptions as before, we get for the single junction ${\rm 
Re}Z_{t,i}(\omega)={\rm Re}Z_t(\omega)/N^2$ \cite{grabert,nazarov}. 
This immediately yields the relation
$\sum_{i}\sqrt{\langle(\Delta\varphi_i)^2\rangle} = 
\sqrt{\langle(\Delta\varphi)^2\rangle}$,
which can be shown to hold also with an arbitrary electromagnetic 
environment $Z(\omega)$ in series with the array. 

If we define the dephasing time $\tau_\varphi$ as the value of $t$ for 
which  $\langle(\Delta\varphi)^2\rangle = (\pi/2)^2$, we obtain for 
zero temperature:
\begin{equation}
\tau_\varphi =
\tau \exp{\left( \frac{\pi^2}{16}\frac{R_K}{R_e}-\gamma\right)}, 
{\rm~~~}(T=0).
\label{tau0}
\end{equation}
Dropping out the small constant terms in Eq.~(\ref{lt}) we can write 
the result at finite temperature in the form:
\begin{equation}
\tau_\varphi \simeq \frac{\pi}{16}\frac{\hbar}{k_BT}\frac{R_K}{R_e}, 
{\rm~~~} (T > 0).
\label{tau}
\end{equation}

\begin{figure}[ht]
\center
\epsfig{file=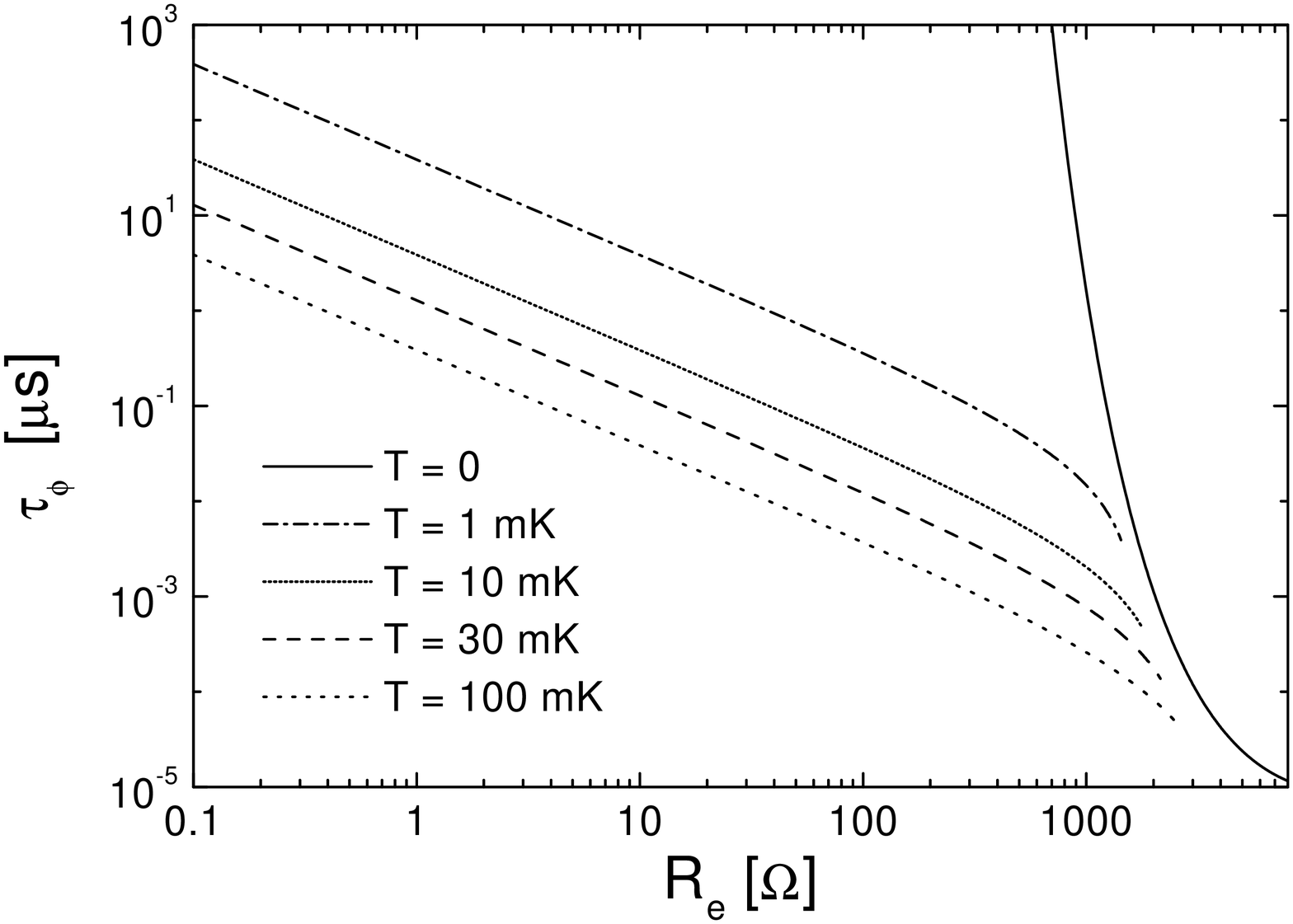, width=85truemm}
\vskip 2mm
\caption{The dephasing time, $\tau_\varphi$, for a three junction 
array as a function of the series resistance $R_e$ of the 
electromagnetic environment. The array is assumed to be homogeneous 
with junction capacitances of $1.0$ fF. The zero temperature curve 
forms a high resistance envelope of the curves corresponding to finite 
temperatures. Finite temperature curves are obtained from 
Eq.~(\ref{lt}) and shown only over their range of validity.}
\label{re}
\end{figure}

For an array with $N=3$ and $C=10^{-15}$ F, and for the environment of 
$R_e = 1$ k$\Omega$ we obtain $\tau_\varphi\approx 1.5$ $\mu$s at zero 
temperature. A resistance of the environment of the order of the free 
space impedance $R_e=Z_0\approx 377$ $\Omega$, yields 
$\tau_\varphi\approx 1.5\cdot10^5$ s. With the same parameters at 
$T=50$ mK the decoherence is very fast: $\tau_\varphi\approx 0.77$ ns 
and $\tau_\varphi\approx 2.1 $ ns for $R_e = 1$ k$\Omega$ and 
$R_e=Z_0$, respectively. Figure \ref{re} shows the dependence of 
$\tau_\varphi$ on the resistance of the environment, $R_e$, for a 
homogeneous three junction array at several different temperatures. 
Also the zero temperature limit (solid line) corresponding to 
Eq.~(\ref{tau0}), is shown. It forms an envelope curve for the finite 
temperature curves calculated from Eq.~(\ref{lt}). It is also seen 
that only in the limit of low environmental resistance, $R_e \leq 1$ 
$\Omega$, one can obtain long decoherence times $\tau_\varphi > 1$ 
$\mu$s at realistic measurement temperatures. Such values of 
$\tau_\varphi^{-1}$ would possibly allow practical quantum logic 
operations and measurements to be performed by fast RF-gate lines and 
by using an RF-SET (radio frequency single electron transistor) as an 
electrometer \cite{rfset}.

\begin{figure}[ht]
\center
\epsfig{file=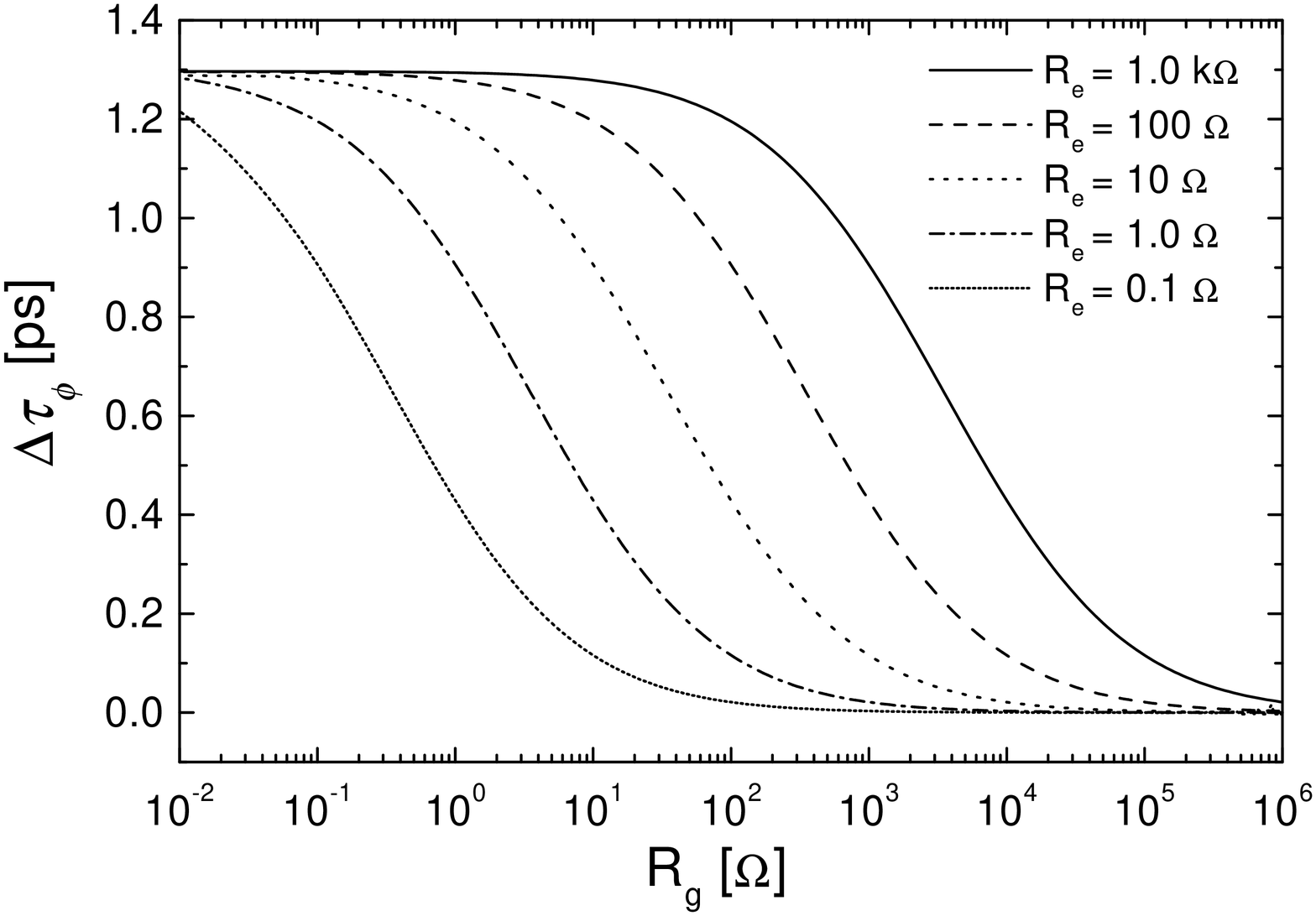,width=85truemm}
\vskip 1mm
\caption{The difference between the dephasing time, $\tau_\varphi$, in 
the three junction Cooper pair pump (Fig.~\ref{schem}b) and the three 
junction array without gate lines (Fig.~\ref{schem}a) as a function of 
resistance, $R_g$, in the gate lines. Capacitances used in the 
calculations are $C_1=C_2=C_3=0.1$ fF and $C_{g1}=C_{g2}=0.01$ fF. 
$T=30$ mK.}
\label{rg}
\end{figure}

In the previous analysis we considered an array of Josephson junctions 
without gates connected capacitively to the islands. These 
capacitances together with the impedance of the gate lines, $Z_{gi}$, 
also affect decoherence and should be taken into account. Therefore a 
quantitative analysis was also applied to the symmetric ($C_i\equiv 
C$, $C_{gi}\equiv C_g$ and $Z_{gi}\equiv R_g$ with all $i$) three 
junction Cooper pair pump which includes gate lines connected to the 
islands, as shown in Fig.~\ref{schem}b. The total impedance seen by 
the array is
\begin{equation}
\label{pumpz}
Z_t(\omega) = \frac{R_e}{1+i\omega\tau\frac{6i\omega C + 3g}{6i\omega 
C + 2g}}\, ,
\end{equation}
where $g^{-1} = R_g + 1/(i\omega C_g)$ is the impedance of the gate 
line and $\tau=R_eC/3$. The explicit expression for $\tau_\varphi$ 
from Re$Z_t(\omega)$ does not assume a simple form but can be 
calculated numerically. Figure \ref{rg} shows the influence of the 
gate lines to the dephasing time as a function of $R_g$ with several 
different values of $R_e$. In the case of disconnected gate lines 
($R_g\rightarrow\infty$) $\tau_{\varphi}$ naturally approaches the 
value of the array without gates, as seen from the figure. In the 
limit of vanishing $R_g$ we also recover the result of an array by 
replacing $C$ by the effective capacitance $C_{\rm 
eff}=C(C+C_g/2)/(C+C_g/3)$ in Eq.~(\ref{ztr}).
 
The influence of dissipation in the gate lines on the dephasing rate 
is counterintuitive at the first sight: gate lines, even resistive 
ones, make the dephasing time longer than in an array without gates 
(Fig.~\ref{rg}). The reason behind this is twofold. Firstly, in our 
estimates we are interested in the fluctuations of the total phase 
difference across the array $\varphi$, not in those of the individual 
phases $\varphi_i$. Because of the series connection with additive 
phase differences, each gate line induces an exactly opposite, i.e. a 
cancelling fluctuation in the neighbouring junctions. Thus the noise 
of the gate resistors does not contribute to the noise in $\varphi$. 
On the other hand, the gate lines decrease the impedance seen by the 
whole array, and this way the noise in the total phase also decreases. 
The longest dephasing time is therefore obtained with non-resistive 
gate lines. Due to the anticorrelated fluctuations in $\varphi_i$ the 
sum rule for $\sqrt{\langle(\Delta\varphi_i)^2\rangle}$, verified 
earlier for an $N$ junction array without gates, does not hold 
anymore. \vskip -3mm


A multijunction Josephson pump provides an interesting testground for 
quantum coherence \cite{omapump} and, on the other hand, it may 
eventually qualify as a metrologically accurate current standard when 
coherence is suppressed by a very dissipative environment 
\cite{rpump}. Here we discuss the three junction pump 
(Fig.~\ref{schem}b) whose characteristics are determined by the two 
energies $E_J$, the Josephson coupling energy, and $E_C$, the charging 
energy. In the ideally coherent adiabatic regime, the phase across the 
array, $\varphi$, is constant and no Landau-Zener band crossing occurs 
\cite{zener}, and the optimum charge transfer through the array 
attains an approximate value (in the lowest order in $E_J/E_C$) 
\cite{omapump}
\begin{equation}
\frac{I}{2ef} \simeq 1-9\frac{E_J}{E_C}\cos(\varphi).
\end{equation}
Here $I$ is the current induced by operating the gates and $f$ is the 
frequency at which the system makes a wind around a degeneracy node of 
the charging energy, i.e.~the frequency of the harmonic gate voltages 
$V_{g1}$. (The two gate voltages are phase-shifted by $\pi/2$.) The 
pumped charge per cycle, $Q=I/f$, is related but not equal to the 
geometric phase (Berry's phase) \cite{omapump,berry,falci} accumulated 
during one cycle along the closed path on the gate plane ($V_{g1}, 
V_{g2}$). Contrary to the pump in the normal state \cite{keller}, the 
coherent adiabatic Josephson pump lacks the ability to pump single 
charges virtually free of errors, and the relative deviations, $\simeq 
-9E_J/E_C\cos(\varphi)$, from the quantized transport are large even 
for very small values of $E_J/E_C$: for example, they can be as large 
as 9\% for $E_J/E_C = 0.01$. Yet, if the gates are operated slow 
enough, which means $f\ll 1/\tau_\varphi$, the $\cos(\varphi)$ term 
averages to $\langle\cos(\varphi)\rangle =0$ during one cycle or 
during the integration time of the measurement, and the pumping 
becomes accurate. Another limit comes from the Landau-Zener band 
crossing, which sets an upper limit for the operation frequency 
$f_{LZ} \simeq E_J^2/(\hbar E_C)$. For typical $E_J = 0.1$ meV and 
$E_C = 1$ meV we obtain $f_{LZ} \approx 10$ GHz. 

\begin{figure}[ht]
\center
\epsfig{file=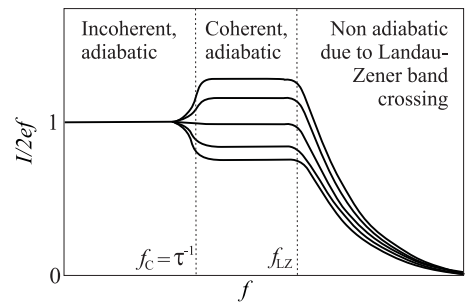,width=60truemm}
\vskip 1mm
\caption{A schematic presentation of the expected behaviour of the 
pumped current in a three junction Cooper pair pump as a function of 
the pumping frequency $f$. Different curves refer to different values 
of the phase difference $\varphi$ across the array.}
\label{mittaus}
\end{figure}

Based on these limitations we expect the following dependence of the 
pump performance at different frequencies (Fig.~\ref{mittaus}). 
\vskip -3mm
\begin{enumerate}
\item\underline{$f<\tau_\varphi^{-1}\equiv f_C$}:~~$I/2ef \simeq 1$ 
because pumping is adiabatic but the phase is undetermined 
$(\:\langle\cos(\varphi)\rangle = 0\:)$. Yet at the lowest frequencies 
the current becomes very small to measure and the accuracy will be 
lost in practise.
\item\underline{$f_C<f<f_{LZ}$}:~~$I/2ef\simeq 
1-9E_J/E_C\cos(\varphi)$, pumping is adiabatic and coherent.
\item\underline{$f>f_{LZ}$}:~~$I/2ef$ decays because the condition for 
no band crossing is lost and charge transport does not follow the 
gating sequence adiabatically.
\end{enumerate}

Since $\tau_\varphi$ is presently expected to fall in the range 
$\tau_\varphi\gg 5$ ns, in a carefully designed experiment, we would 
have $f_C \ll 200$ MHz, yielding a clear separation of the three 
pumping regimes. In particular, if the decoherence time of a squbit 
and thus also $\tau_\varphi$ turns out to be 
of the order of $1$ $\mu$s, which would allow quantum computation by 
Josephson qubits in this respect, $\tau_\varphi^{-1}$ would give an 
experimentally convenient crossover frequency in the MHz range. 

Our estimations of the decoherence time also bring up the issue of 
using so-called quasiparticle traps in single Cooper pair boxes and 
transistors. The parity effect has been observed over the years in 
several experiments \cite{tuominen,saclay}. It is the manifestation of 
pure Cooper pair effects (not necessarily coherent) in small 
superconducting islands. Up to now, using aluminium structures, parity 
effect manifested by a 2e-periodic gate modulation, can be observed 
reliably only in Josephson junction circuits which are embedded in a 
resistive environment. Our estimates of $\tau_\varphi$ now set the 
limit of how dissipative the quasiparticle traps can be, in order not 
to destroy the coherent state of the qubits too fast.  

To perform an experiment of Fig.~\ref{mittaus} with distinct regimes 
one has to have a setup with long enough dephasing time 
$\tau_\varphi\geq 10$ ns. This means that the on-chip resistances 
should be very low, which limits the use of quasiparticle traps. It 
has already been shown by experiment that with high enough resistances 
in the biasing circuit the pumped current becomes accurate 
\cite{rpump2}. Combined with the fact that the parity effect, i.e.~the 
quasiparticle free Cooper pair effect is very difficult to observe 
without quasiparticle traps, one needs to seek alternative measurement 
schemes of the pump. One way to avoid decoherence induced by a 
quasiparticle current is to fabricate an on-chip loop of a gated array 
terminated by an on-chip SET-electrometer \cite{keller}. The 
electrometer could then be used to measure the number of Cooper pairs 
pumped through the array into it. However, it turns out that in this 
case the pumped current is accurate ($I = 2ef$) due to charge 
conservation implied by the terminating classical capacitance. 

Our suggestion is to realise the experiment shown in 
Fig.~\ref{mittaus} by using a closed superconducting (phase) biasing 
circuit on the chip with an inductance in series. This way the real 
part of the impedance of the series environment, ${\rm 
Re}Z_t(\omega)$, vanishes. Further, gate lines do not induce any extra 
decoherence to the system, as shown before, and can therefore be as 
resistive as needed to filter the feed lines. Thus the major source of 
decoherence is the resistive impedance of the quasiparticle traps if 
needed. The pumped current can be measured by a SQUID ammeter 
inductively connected to the coil in the biasing circuit. This kind of 
a setup might give a low enough decoherence rate.

In conclusion, we have presented a method to quantitatively estimate 
the decoherence time, $\tau_\varphi$, due to dissipative 
electromagnetic environment in circuits consisting of small Josephson 
junctions. This method allows us, among other things, to discuss the 
suitability of the system in consideration as a quantum bit. We also 
suggest a direct measurement of $\tau_\varphi^{-1}$ as a crossover 
pumping frequency between coherent and incoherent pumping in the 
single Cooper pair pump.

{\bf Acknowledgments:}
This work has been supported by the Academy of Finland under the 
Finnish Centre of Excellence Programme 2000-2005 (Project No. 44875, 
Nuclear and Condensed Matter Programme at JYFL) and EU (contract 
IST-1999-10673). We thank Klavs Hansen and Matias Aunola for many 
useful discussions.



\vskip -5mm
\begin{thebibliography}{99}
\vspace{-13mm}
\bibitem{nakamura} Y. Nakamura, Yu.A. Pashkin, and J.S. Tsai, Nature 
{\bf 396}, 786 (1999);
Y. Nakamura, {\sl Private communication}.
\bibitem{averin} D.V. Averin, Solid State Commun. {\bf 105}, 659 
(1998).
\bibitem{schon} Yu. Makhlin, G. Sch\"on, A. Shnirman, Nature {\bf 
398}, 305 (1999); G. Sch\"on, A. Shnirman, Yu. Makhlin, 
cond-mat/9811029.
\bibitem{cottet} A. Cottet, A. Steinbach, P. Joyez, D. Vion, H. Pothier,
D. Esteve, and M.E. Huber, in the proceedings of {\sl Macroscopic Quantum 
Coherence and Computing}, Naples, 14-17 June (2000), in print.
\bibitem{omapump} J.P. Pekola, J.J. Toppari, M. Aunola, M.T. 
Savolainen, and D.V. Averin, Phys. Rev. B {\bf 60}, R9931 (1999); M. 
Aunola, J.J. Toppari, and J.P. Pekola, Phys. Rev. B {\bf 62}, 1296 
(2000).
\bibitem{joyez} P. Joyez, {\sl Le Transistor a une Paire de Cooper: Un 
Systeme Quantique Macroscopique}, Academic Dissertation for the Degree 
of Doctor of Philosophy, University of Paris 6 (1995). 
\bibitem{grabert} H. Grabert, G-L. Ingold, M.H. Devoret, D. Esteve, H. 
Pothier, and C. Urbina, Z. Phys. B {\bf 84}, 143 (1991).
\bibitem{nazarov} G.-L. Ingold, and Yu.V. Nazarov, in {\sl Single 
Charge Tunnelling, Coulomb Blockade Phenomena in Nanostuctures}, 
edited by H. Grabert, and M.L. Devoret (Plenum Press, New York, 1992). 
\bibitem{jt} D.S. Golubev, and A.D. Zaikin, in {\sl Quantum Physics at 
Mesoscopic Scale}, eds. D.C. Glattli, M. Sanquer and J. Tran Thanh Van 
(Frontieres, 1999).
\bibitem{jt2} Yu.V. Nazarov, Sov. Phys. JETP {\bf 68}, 561 (1989); M. 
H. Devoret, D. Esteve, H. Grabert, G.-L. Ingold, H. Pothier, and C. 
Urbina, Phys. Rev. Lett. {\bf 64}, 1824 (1990).
\bibitem{rfset} R. J. Schoelkopf, P. Wahlgren, A. A. Kozhevnikov, P. 
Delsing, and D. E. Prober, Science {\bf 280}, 1238 (1998).  
\bibitem{rpump} A. B. Zorin, S. V. Lotkhov, H. Zangerle, and J. 
Niemeyer, J. Appl. Phys. {\bf 88}, 2665 (2000); S. V. Lotkhov, S. A. 
Bogoslovsky, A. B. Zorin, J. Niemeyer, cond-mat/0001206. 
\bibitem{zener} See, e.g., J.M. Ziman, {\sl Principles
of the theory of solids}, (Cambridge University Press, Cambridge, 
1964).
\bibitem{berry} M.W. Berry, Proc. R. Soc. London, Ser. A {\bf 392}, 45 
(1984).
\bibitem{falci} G. Falci, R. Fazio, G.M. Palma, J. Siewert, V. Vedral, 
Nature {\bf 407}, 355 (2000).
\bibitem{keller} M.W. Keller, J.M. Martinis, N.N. Zimmerman, A.H. 
Steinbach, Appl. Phys. Lett. {\bf 69}, 1804 (1996); M.W. Keller, J.M. 
Martinis, and R.L. Kautz, Phys. Rev. Lett. {\bf 80}, 4530 (1998); M.W. 
Keller, A.L. Eichenberger, J.M. Martinis, and N.M. Zimmerman, Science 
{\bf 285}, 1716 (1999).
\bibitem{tuominen} M. T. Tuominen, J. M. Hergenrother, T. S. Tighe, 
and M. Tinkham,
Phys. Rev. Lett {\bf 69}, 1997 (1992); M. T. Tuominen, J. M. 
Hergenrother, T. S. Tighe, and M. Tinkham, Phys. Rev. B {\bf 47}, 
11599 (1993).
\bibitem{saclay} P. Lafarge, P. Joyez, D. Esteve, C. Urbina, and M. H. 
Devoret, Phys. Rev. Lett. {\bf 70}, 994 (1993). 
\bibitem{rpump2} A.B. Zorin, S.A. Bogoslovsky, S.V. Lotkhov, J. 
Niemeyer, 
in proceedings of Macroscopic Quantum Coherence and Quantum Computing 
(MQC2) Naples, edited by D. Averin, B. Ruggiero, and P. Silvestrini, 
(Plenum Publishers, New York, 2000).
\end{thebibliography}
\end{document}